\documentclass[12pt]{extarticle}
\usepackage{setspace}

\usepackage{setspace}
\usepackage[T1]{fontenc}
\usepackage{times}
\usepackage{palatino} 
\usepackage{lmodern}
\usepackage[latin1]{inputenc}
\usepackage{epsfig}
\usepackage[english]{babel}
\usepackage{color}
\usepackage{graphicx}
\usepackage{dcolumn}
\usepackage{moreverb}
\usepackage{amsmath,amssymb,amsfonts}
\usepackage[all]{xy}

\usepackage[svgnames]{xcolor}
\usepackage{tikz}

\def\nudge{.5}

\tikzset{axis/.style={ultra thick, Red!75!black, -latex, shorten <=-\nudge cm, shorten >=-2*\nudge cm}}
\tikzset{line/.style={thick,Green}}

\begin{document}
\numberwithin{equation}{section}
\newcommand{\boxedeqn}[1]{%
  \[\fbox{%
      \addtolength{\linewidth}{-2\fboxsep}%
      \addtolength{\linewidth}{-2\fboxrule}%
      \begin{minipage}{\linewidth}%
      \begin{equation}#1\end{equation}%
      \end{minipage}%
    }\]%
}


\newsavebox{\fmbox}
\newenvironment{fmpage}[1]
     {\begin{lrbox}{\fmbox}\begin{minipage}{#1}}
     {\end{minipage}\end{lrbox}\fbox{\usebox{\fmbox}}}

\raggedbottom
\onecolumn

\parindent 8pt
\parskip 10pt
\baselineskip 16pt
\noindent\title*{{\LARGE{\textbf{Quadratic algebra for superintegrable monopole system in a Taub-NUT space}}}}
\newline
\newline
\newline
Md Fazlul Hoque$^a$, Ian Marquette$^a$ and Yao-Zhong Zhang$^{a,b}$
\newline
\newline
$^a$School of Mathematics and Physics, The University of Queensland, Brisbane, QLD 4072, Australia
\newline
\newline
$^{b}$CAS Key Laboratory of Theoretical Physics, Institute of Theoretical Physics, Chinese Academy of Science, Beijing 100190, China
\newline
\newline
E-mail: m.hoque@uq.edu.au; i.marquette@uq.edu.au; yzz@maths.uq.edu.au
\newline
\newline
\begin{abstract}
We introduce a Hartmann system in the generalized Taub-NUT space with Abelian monopole interaction. This quantum system includes well known Kaluza-Klein monopole and MIC-Zwanziger monopole as special cases.  It is shown that the corresponding Schr\"{o}dinger equation of the Hamiltonian is separable in both spherical and parabolic coordinates. We obtain the integrals of motion of this superintegrable model and construct the quadratic algebra and Casimir operator. This algebra can be realized in terms of a deformed oscillator algebra and has finite dimensional unitary representations (unirreps) which provide energy spectra of the system. This result coincides with the physical spectra obtained from the separation of variables.
\end{abstract}

\section{Introduction}
Dirac first explored the existence of monopoles in the quantum mechanical interest and the quantization of electric charge \cite{Dir1}. Later the Kepler problems involving additional magnetic monopole interaction was independently discovered by McIntosh and Cisneros \cite{Mci1} and  Zwanziger \cite{Zwa1} which is known as MICZ-Kepler problem. The MICZ-Kepler problem discusses the existence of the Runge-Lenz vector in addition to the angular momentum vector and a large dynamical symmetry $so(4)$ algebra \cite{Bar1,Jac1}. The generalized Dirac monopole exhibits a hidden dynamical algebra \cite{Men1}. The MICZ-Kepler problems have been generalized using many approaches to higher dimensions \cite{ Men1, Yan1,Mad1, Ner1, Mad4}. These monopole systems are separable in hyperspherical, spheroidal and parabolic coordinates \cite{ Men1, Yan1, Mad1, Ner1, Mad4}. 

One important class of monopole models is Kaluza-Klein monopoles. Kaluza and Klein  introduced a five-dimensional theory with one dimension curled up to form a circle in the context of unification theory \cite{Kak1, Kak2}. The complete algebraic description of Kaluza-Klein monopole allows a dynamical symmetry of the quantum motions \cite{Grs1, Gib1, Feh1, Cor1}. Models in space with Taub-NUT (Taub-Newman-Unit-Tambrino) metric have attracted much attention because the geodesic of the Taub-NUT metric describes appropriately the motion of well-separated monopole-monopole interactions ( see e.g. \cite{Grs1, Gib1, Feh1, Cor1, Man2, Ati1, Gro1, Cot1, Mar1}). This Taub-NUT metric is well known to admit the Kepler-type symmetry  and provides non-trivial generalization of the Kepler problems. Iwai and his collaborators published a series of papers  \cite{Iwa5, Iwa2, Iwa1, Iwa6} about the reduction system to admit a Kepler and harmonic oscillator type symmetry via generalized Taub-NUT metric. The generalized MICZ-Kepler problems  \cite{Mad5} represent the intrinsic Smorodinsky-Winternitz system \cite{Fri1,Eva2} with monopole in 3D Euclidean space. Supersymmetry could be constructed in the generalized MICZ-Kepler system \cite{Ran1}. The MICZ-Kepler problem was also considered in $S^3$ \cite{Gri1}.

Quadratic algebra is a useful tool to obtain energy spectrum of superintegrable systems in the viewpoint of classical and quantum mechanics \cite{Gra1}. General quadratic algebras involving three generators generated by second-order integrals of motion and their realizations in terms of deformed oscillator algebra have been investigated in \cite{Das1}. Most of the applications of the quadratic algebra and representation theory have been on systems with scalar potential interactions \cite{Kal1, Kal2, Das2, Mil1, Isa1, Gen1, FH1,FH2}. In this paper we introduce a new superintegrable system in a Taub-NUT space with Abelian monopole interaction. 

The contents of this paper are organized as follows: Section 2 introduces a new Kepler monopole system in a Taub-NUT space which includes Kaluza-Klein and MICZ monopole as special cases. It is remarked that this system covers a class of dynamical systems of interest. In section 3, the Schrodinger equation of the Hamiltonian in the Taub-NUT space is solved in both spherical and parabolic coordinates. In section 4, we construct second-order integrals of motion which show the superintegrability of the model in the parabolic coordinates. We obtain quadratic algebra and Casimir operator generated by integrals, and realize these algebras in terms of deformed oscillator algebra. This enables us to obtain the energy spectra of the system algebraically. Finally in section 5, we discuss the results and some open problems.

\section{Kepler monopole system}
Let us consider the generalized Taub-NUT metric in $\mathbb{R}^3$
\begin{eqnarray}
ds^2=f(r)d\textbf{r}^2+g(r)(d\psi+\textbf{A}. d\textbf{r} )^2,\label{mc1}
\end{eqnarray}
where 
\begin{eqnarray}
&& f(r)=\frac{a}{r}+b, \qquad g(r)=\frac{r(a+br)}{1+c_1 r+dr^2},\label{fg1}
\\&&
A_1=\frac{-y}{r(r+z)}, \quad A_2=\frac{x}{r(r+z)}, \quad A_3=0,
\end{eqnarray}
$r=\sqrt{x^2+y^2+z^2}$ and the three dimensional Euclidean line element $d\textbf{r}^2=dx^2+dy^2+dz^3$, $a$, $b$, $c_1$, $d$ are constants. Here $\psi$ is the additional angular variable which describes the relative phase and its coordinate is cyclic with period $4\pi$ \cite{ Grs1, Cor1}. The functions $f(r)$ and $g(r)$ in the metric represent gravitational effects and $A_i$ is identified the monopole interaction. 

We consider the Hamiltonian system associated with (\ref{mc1}) 
\begin{eqnarray}
H=\frac{1}{2}\left[\frac{1}{f(r)}\left\{p^2+\frac{c_0}{2r}+\frac{c_2}{2r(r+z)}+\frac{c_3}{2r(r-z)}+c_4\right\}+\frac{Q^2}{g(r)}\right],\label{k1}
\end{eqnarray}
where $c_0$, $c_2$, $c_3$, $c_4$ are constants and the operators
\begin{eqnarray}
p_i=-i(\partial_i-i A_i Q), \quad Q=-i\partial_\psi
\end{eqnarray}
satisfying the following commutation relations
\begin{eqnarray}
[p_i,p_j]=i\epsilon_{ijk}B_k Q, \quad [p_i,Q]=0,\quad \textbf{B}=\frac{\textbf{r}}{r^3}.
\end{eqnarray} 
The system with Hamiltonian (\ref{k1}) is generalized Hartmann system \cite{Hartm1} in a curved Taub-NUT space with abelian monopole interaction. The Hartmann system is a deformed Coulomb interaction in 3D Euclidean space. This new system (\ref{k1}) is referred to as Kepler monopole system. It contains the Kaluza-Klein \cite{Gib1, Mar1} and MICZ monopoles \cite{Mci1, Zwa1} as special cases:  it is Kaluza-Klein monopole system when $a=1$, $b=1$, $c_1=2$, $d=1$ and MICZ monopole system when $a=0$, $b=1$, $c_1=-2$, $d=1$, $c_2=0$, $c_3=0$.

The Kepler monopole system allows the following suitable total angular momentum operator $\textbf{L}$ and the Runge-Lenz operator \textbf{M} which can be constructed into the form
\begin{eqnarray}
\textbf{L}=\textbf{r}\times \textbf{p}-\frac{\textbf{r}}{r}Q, \quad \textbf{M}=\frac{1}{2}(\textbf{p}\times \textbf{L}-\textbf{L}\times \textbf{p})-\frac{\textbf{r}}{r}(aH-\frac{c_1}{2} Q^2).
\end{eqnarray}
The operators $L$ and $M$ commute with the Kepler monopole system (\ref{k1}) when $c_0=c_2=c_3=c_4=0$ and verify its maximally superintegrability.
These operators close to an $o(4)$ or $o(3,1)$ dynamical symmetry algebra in the quantum state with fixed energy and eigenvalue of operator $Q$:
\begin{eqnarray}
[L_i, L_j]=i\epsilon_{ijk}L_k,\quad [L_i, M_j]=i\epsilon_{ijk}M_k, \quad [M_i, M_j]=i\epsilon_{ijk}L_k(\frac{c_1 Q^2}{2}-aH).
\end{eqnarray}
It is $o(4)$ algebra for $\frac{c_1 Q^2}{2}-aH>0$  and $o(3,1)$ algebra for $\frac{c_1 Q^2}{2}-aH<0$.

In the next section, we examine model (\ref{k1}) in the spherical and parabolic coordinate systems for separation of variables.

\section{Separation of variables}
The Hamiltonian (\ref{k1}) with monopole interaction is multiseparable and allows the separation of variables for the corresponding Schr\"{o}dinger equations in spherical and parabolic coordinates.

\subsection{Spherical coordinates}
Let us consider the spherical coordinates 
\begin{eqnarray}
&&x=r \sin\theta\cos\phi,\quad y=r\sin\theta\sin\phi,\quad z=r\cos\theta,
\end{eqnarray}
where $r>0$, $0\leq\theta\leq\pi$ and $0\leq \phi\leq 2\pi$.
In terms of these coordinates, the Taub-NUT metric (\ref{mc1}) takes on the form 
\begin{eqnarray}
ds^2=f(r)(dr^2+r^2d\theta^2+r^2\sin^2\theta d\phi^2)+g(r)(d\psi+\cos\theta d\phi)^2,\label{mc2}
\end{eqnarray}
\begin{eqnarray}
A_1=-\frac{1}{r}\tan\frac{\theta}{2}\sin\phi,\quad A_2=\frac{1}{r}\tan\frac{\theta}{2}\cos\phi,\quad A_3=0,
\end{eqnarray}
and the Schrodinger equation $H\Psi=E\Psi$ of the model (\ref{k1}) takes the following form 
\begin{eqnarray}
&&\frac{-r}{2(a+br))}\left[\frac{\partial^2}{\partial r^2}+\frac{2}{r}\frac{\partial}{\partial r}-\frac{c_0}{2r}-\frac{c_2}{4r^2\cos^2\frac{\theta}{2}}-\frac{c_3}{4r^2\sin^2\frac{\theta}{2}}-c_4\right.\nonumber\\&&\left.+\frac{1}{r^2}\left(\frac{\partial^2}{\partial\theta^2}+\cot\theta\frac{\partial}{\partial\theta}+\frac{1}{\sin^2\theta}\frac{\partial^2}{\partial\phi^2}\right)+\left(\frac{1}{r^2\cos^2\frac{\theta}{2}}+\frac{c_1}{r}+d\right)\frac{\partial^2}{\partial\psi^2}\right.\nonumber\\&&\left.-\frac{1}{r^2\cos^2\frac{\theta}{2}}\frac{\partial}{\partial\phi}\frac{\partial}{\partial\psi}   \right]\Psi(r,\theta,\phi,\psi) =E\Psi(r,\theta,\phi,\psi).\label{kp2}
\end{eqnarray}
For the separation of (\ref{kp2}), the ansatz
\begin{eqnarray}
\Psi(r,\theta,\phi,\psi)=R(r)\Theta(\theta)e^{i(\nu_1\phi+\nu_2\psi)},
\end{eqnarray}
leads readily to the following radial and angular ordinary differential equations   
\begin{eqnarray}
&&\left[\frac{d^2}{d r^2}+\frac{2}{r}\frac{d}{d r}+\alpha+\frac{\beta}{r}-\frac{k_1}{r^2}\right ]R(r)=0,\label{kp3}
\\
&&\left[\frac{d^2}{d\theta^2}+\cot\theta\frac{d}{d\theta}+\left\{k_1-\frac{c_2+(\nu_1-2\nu_2)^2}{2(1+\cos\theta)}-\frac{c_3+\nu_1^2}{2(1-\cos\theta)}\right\}\right]\Theta(\theta)=0,\nonumber\\&&\label{kp4}
\end{eqnarray}
where $\alpha=2b E-d \nu_2^2-c_4$, $\beta=2a E-c_1\nu_1^2-\frac{c_0}{2}$ and $k_1$ is separable constant.
We now turn to (\ref{kp4}), which can be converted, by setting $z=\cos\theta$ and $\Theta(z)=(1+z)^{a}(1-z)^{b} Z(z)$, to
\begin{eqnarray}
&&(1-z^2)Z''(z)+\{2a-2b-(2a+2b+2)z\}Z'(z)\nonumber \\&&\qquad +\{k_2-(a+b)(a+b+1)\}Z(z)=0,\label{pr5}
\end{eqnarray}
where $2a=\delta_{1}+\nu_1$, $2b=\delta_{2}+\nu_1$ and
\begin{eqnarray}
\delta_{1}=\sqrt{c_2+(\nu_1-2\nu_2)^2}-\nu_1,\quad \delta_{2}=\sqrt{c_3+\nu_1^2}-\nu_1. \label{pr1}
 \end{eqnarray}
Comparing (\ref{pr5}) with the Jacobi differential equation  
\begin{equation}
(1-x^{2})y''+\{\beta_1-\alpha_1-(\alpha_1+\beta_1+2)x\}y'+\lambda(\lambda+\alpha_1+\beta_1+1)y=0,\label{Jd1}
\end{equation}
we obtain the  separation constant 
\begin{equation}
k_1=(l+\frac{\delta_{1}+\delta_{2}}{2})(l+\frac{\delta_{1}+\delta_{2}}{2}+1),\label{pr2} 
\end{equation}
where $l=\lambda+\nu_1$. 
Hence solutions of (\ref{kp4}) are given in terms of the Jacobi polynomials as
\begin{eqnarray}
\Theta(\theta)&\equiv &\Theta_{l \nu_1}(\theta; \delta_{1}, \delta_{2})
= F_{l \nu_1}(\delta_{1}, \delta_{2})(1+\cos\theta)^{\frac{(\delta_{1}+\nu_1)}{2}}(1-\cos\theta)^{\frac{(\delta_{2}+\nu_1)}{2}}\nonumber\\&&\quad\times P^{(\delta_{2}+\nu_1, \delta_{1}+\nu_1)}_{l-\nu_1}(\cos\theta),\label{jp1}
\end{eqnarray}
where $P^{(\alpha, \beta)}_{\lambda}$ denotes Jacobi polynomial, $F_{l \nu_1}(\delta_{1}, \delta_{2})$ is the normalized constant and $l\in \mathbb{N}$. 

Let us now turn to the radial equation (\ref{kp3}), which can be converted, by setting  
 $z=\varepsilon r$, $R(z)=z^{l+\frac{\delta_{1}+\delta_{2}}{2}} e^{-\frac{z}{2}}R_1(z)$ and $\alpha =\frac{-\varepsilon^2}{4}$, to
\begin{equation}
z\frac{d^2R_1(z)}{dz^2}+\{(2l+\delta_{1}+\delta_{2}+2)-z\}\frac{dR_1(z)}{dz}-(\frac{\delta_{1}+\delta_{2}}{2}+l+1)-\frac{\beta}{\varepsilon})R_1(z)=0.\label{an6}
\end{equation}
Set  
\begin{eqnarray}
n=\frac{\beta}{\varepsilon}-\frac{\delta_{1}+\delta_{2}}{2}.\label{an7}
\end{eqnarray}
Then (\ref{an6}) can be expressed as
\begin{equation}
z\frac{d^2R_1(z)}{dz^2}+\{(2l+\delta_{1}+\delta_{2}+2)-z\}\frac{dR_1(z)}{dz}-(-n+l+1)R_1(z)=0.\label{an8}
\end{equation}
This is the confluent hypergeometric equation. Hence we can write the solution of (\ref{kp3}) in terms of the confluent hypergeometric function as  
\begin{eqnarray}
&R(r)&\equiv R_{nl}(r;\delta_{1}, \delta_{2})=F_{nl}(\delta_{1},\delta_{2})(\varepsilon r)^{l+\frac{\delta_{1}+\delta_{2}}{2}} e^{\frac{-\varepsilon r}{2}}\nonumber\\&&
\quad \times {}_1 F_1(-n+l+1, 2l+\delta_{1}+\delta_{2}+2; \varepsilon r),\label{an9}
\end{eqnarray}
where $F_{nl}(\delta_{1},\delta_{2})$ is the normalized constant.
In order to have a discrete spectrum the parameter $n$ needs to be positive integer. From (\ref{an7}) we have  
\begin{equation}
\varepsilon=\frac{\beta}{(n+\frac{\delta_{1}+\delta_{2}}{2})}
\end{equation} and hence the energy spectrum is given by
\begin{equation}
\frac{2aE-c_1\nu_2^2-\frac{c_0}{2} }{2\sqrt{c_4-2b E+d\nu_2^2}}=n+\frac{\delta_1+\delta_2}{2},\qquad n=1, 2, 3,\dots\label{en2}
\end{equation}

\subsection{Parabolic Coordinates }
The parabolic coordinate system has the form
\begin{eqnarray}
&&x=\sqrt{\xi\eta}\cos\phi,\quad y=\sqrt{\xi\eta}\sin\phi,\\&&
z=\frac{1}{2}(\xi-\eta),\quad r=\frac{1}{2}(\xi+\eta)
\end{eqnarray}
with $\xi, \eta >0$ and $0\leq\phi\leq 2\pi$. 
In terms of the coordinates, the Taub-NUT metric (\ref{mc1}) takes the form
\begin{eqnarray}
&&ds^2=f(r)\left[(\xi+\eta)(d\xi^2+d\eta^2)+\xi\eta d\phi^2\right]+g(r)\left[ d\psi+\left(1-\frac{\xi-\eta}{\xi+\eta} d\phi\right) \right]^2,\nonumber\\&&\label{mc3}
\end{eqnarray}
\begin{eqnarray}
A_1=\frac{-2 \sqrt{\eta}}{\sqrt{\xi}}\frac{\sin\phi}{\xi+\eta}, \quad A_2=\frac{-2 \sqrt{\eta}}{\sqrt{\xi}}\frac{\cos\phi}{\xi+\eta}, \quad A_3=0
\end{eqnarray}
and the Schrodinger equation of the system (\ref{k1}) is
\begin{eqnarray}
&&\frac{-1}{2\{2a+b(\xi+\eta)\}}\left[4\xi\frac{\partial^2}{\partial\xi^2}+ 4\eta\frac{\partial^2}{\partial\eta^2}+4\frac{\partial}{\partial\xi} +4\frac{\partial}{\partial\eta}-c_0-\frac{c_2}{\xi}-\frac{c_3}{\eta}\right.\nonumber\\&&\left.-c_4(\xi+\eta)+\frac{\xi+\eta}{\xi\eta}\frac{\partial^2}{\partial\phi^2}-\frac{4}{\xi}\frac{\partial}{\partial\phi}\frac{\partial}{\partial\psi}+\left\{\frac{4}{\xi}+c_1+\frac{d}{4}(\xi+\eta)\right\}\frac{\partial^2}{\partial\psi^2}\right]\nonumber\\&&\times \Psi(\xi,\eta,\phi,\psi)=E\Psi(\xi,\eta,\phi,\psi).\label{kp1}
\end{eqnarray}
By making the Ansatz,
\begin{eqnarray}
\Psi(\xi,\eta,\phi,\psi)=\Theta_1(\xi)\Theta_2(\eta)e^{i(\nu_1\phi+\nu_2\psi)},
\end{eqnarray}
(\ref{kp1}) becomes  
\begin{eqnarray}
&&\left[\partial_{\xi}(\xi\partial_{\xi})+\frac{\alpha}{4}\xi+\frac{\beta}{4}-\frac{c_2+(\nu_1-2\nu_2)^2}{4\xi}\right ]\Theta_1(\xi)=k_2 \Theta_1(\xi),\label{fk4}
\\
&&\left[\partial_{\eta}(\eta\partial_{\eta})+\frac{\alpha}{4}\eta+\frac{\beta}{4}-\frac{c_3+\nu_1^2}{4\eta}\right]\Theta_2(\eta)=-k_2 \Theta_2(\eta),\label{fk5}
\end{eqnarray}
where $\alpha=2b E-d\nu_2^2-c_4$, $\beta=2a E-c_1\nu_2^2-\frac{c_0}{2}$ and $k_2$ is separable constant. Putting $z_{1}=\varepsilon \xi $ in (\ref{fk4}), $z_{2}=\varepsilon \eta $ in (\ref{fk5}) and $\Theta_i(z_i)=z_i^\frac{\delta_i+\nu_1}{2}e^{-\frac{z_i}{2}}F_i(z_i)$, $\alpha=-\varepsilon^2$, these two equations represent
\begin{equation}
z_{i}\frac{d^2F_{i}(z_i)}{dz^2_{i}}+\{(\delta_{i}+\nu_1+1)-z_{i})\}\frac{dF_{i}(z_i)}{dz_{i}}-(\frac{\delta_{i}+\nu_1+1}{2}-\frac{\beta}{4\varepsilon}+\frac{k_i}{\varepsilon})F_{i}(z_i)=0,\label{fk6}
\end{equation}
where $i=1,2$, $k_2=-k_1$ and 
\begin{eqnarray}
\delta_1=\sqrt{c_2+(\nu_1-2\nu_2)^2}-\nu_1,\quad \delta_2=\sqrt{c_3+\nu_1^2}-\nu_1.\label{dt1}
\end{eqnarray}
Let us now denote 
\begin{eqnarray}
n_{i}=-\frac{1}{2}(\delta_{i}+\nu_1+1)+\frac{\beta}{4\varepsilon}-\frac{k_i}{\varepsilon}, \quad i=1,2.\label{fk8}
\end{eqnarray}
Then (\ref{fk6}) can be identified with the Laguerre differential equation. Thus we have the normalized wave function
 \begin{eqnarray} 
\Psi(\xi,\eta,\phi,\psi)&=&U_{n_{1}n_{2}\nu_1}(\xi,\eta,\phi,\psi; \delta_{1},\delta_{2})\nonumber\\&=&\frac{\hbar \varepsilon^2}{\sqrt{-8c_{0}}}f_{n_{1}\nu_1}(\xi;\delta_{1})f_{n_{2}\nu_1}(\eta;\delta_{2})\frac{e^{i(\nu_1\phi+\nu_2\psi)}}{\sqrt{2\pi}}, 
\end{eqnarray}
where
\begin{eqnarray*}
 &f_{n_{i} \nu_1}(t_{i};\delta_{i})& \equiv f_{i}(t_{i})=\frac{1}{\Gamma(\nu_1+\delta_{i}+1)}\sqrt{\frac{\Gamma(n_{i}+\nu_1+\delta_{i}+1)}{ n_{i}!}} 
 (\varepsilon t_{i})^{(\nu_1+\delta_{i})/2} e^{-\varepsilon t_{i}/2}\nonumber\\&& \times {}_1F_{1}(-n_{i}, \nu_1+\delta_{i}+1; \varepsilon t_{i}),\label{fk7}   
\end{eqnarray*}
 $i=1, 2$ and $ t_{1}\equiv\xi , t_{2}\equiv\eta $. We look for the discrete spectrum and thus $n_{1}$ and $n_{2}$ are both positive integers. The expression for the energy of the system in terms of $n_{1}$ and $n_{2}$  can be found by using $\alpha=-\varepsilon^2$ in (\ref{fk8}) to be 
\begin{equation}
\frac{2a E-c_1\nu_2^2-\frac{c_0}{2}}{2\sqrt{c_4-2bE+d\nu_2^2}}=n_1+n_2+\frac{\delta_1+\delta_2}{2}+\nu_1+1.\label{en3}
\end{equation}
We can relate the quantum numbers in (\ref{en2}) and (\ref{en3}) by the following relation
\begin{eqnarray}
n_{1}+n_{2}+\nu_1=n-1,
\end{eqnarray}
where $n_{1}, n_{2}=0, 1, 2,\dots.$

\section{Kepler monopole in Taub-NUT space and algebra structure}
In this section, we construct integrals of motion for the superintegrable monopole system (\ref{k1}), their quadratic algebra and Casimir operator. The realization of the algebra in terms of deformed oscillator algebra which generate a finite dimensional unitary representation (unirrep) to degenerate energy spectrum of the model (\ref{k1}) is presented.

\subsection{Integrals of motion and quadratic algebra}
The Hamiltonian system (\ref{k1}) has the following algebraically independent integrals of motion 
\begin{eqnarray}
&&A=L^2+\frac{c_3 \xi}{4\eta}+\frac{c_2\eta}{4\xi}, \label{k2}
\\&&
B=M_3+\frac{\xi-\eta}{\xi+\eta}\left\{\frac{c_3\xi^2-c_2\eta^2}{2\xi^2\eta-2\xi\eta^2}+\frac{c_0}{4}\right\}\label{k3},
\\&& L_3,
\end{eqnarray}
where $L_3=\sqrt{\xi\eta}\cos\phi p_2-\sqrt{\xi\eta}\sin\phi p_1-\frac{\xi+\eta}{\xi-\eta}Q$ and $M_3=\frac{1}{2}\{(p_1L_2-p_2L_1)-(L_1p_2-L_2p_1)\}-\frac{z}{r}(aH-\frac{c_1}{2}Q^2).$
The integral of motion $A$ is associated with the separation of variables in spherical coordinates and $B$ is associated with the separation of variables in parabolic coordinates system. The Hamiltonian (2.4) is minimally superintegable as it allows five integrals of motion including $H$ and the superintegrability can be verified by proving the commutation relations
\begin{eqnarray}
&[A, L_3]=0=[A,H], \quad [B, L_3]=0=[B,H],&\\&[A,Q]=0=[B,Q], \quad [H, Q]=0=[H, L_3], \\&[Q,L_3]=0.&
\end{eqnarray}
For convenience we present a diagram of the above commutation relations
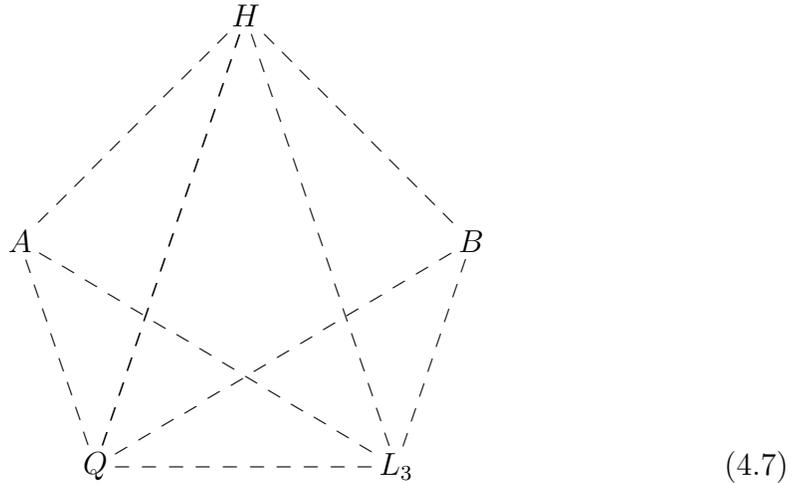
\begin{eqnarray}
\begin{xy}
(10,0)*+{Q}="f"; (50,0)*+{L_{3}}="k"; (0,30)*+{A}="a"; (60,30)*+{B}="b"; (30,60)*+{H}="h";  
"f";"k"**\dir{--}; 
"h";"b"**\dir{--};
"h";"f"**\dir{--};
"h";"a"**\dir{--};
"h";"k"**\dir{--};
"h";"f"**\dir{--};
"a";"f"**\dir{--}; 
"a";"k"**\dir{--};
"b";"f"**\dir{--};
"b";"k"**\dir{--};
\end{xy}
\end{eqnarray}
The diagram shows that $Q$ and $L_3$ are central elements. 

We now construct a new integral of motion $C$ of the system from (\ref{k2}) and (\ref{k3}) via commutator 
\begin{eqnarray}
[A,B]=C.
\end{eqnarray}
Here $C$ is a cubic function of momenta. By direct computation we can show that the integrals of motion $A$, $B$  and central elements $H$, $Q$, $L_3$ satisfy the following quadratic algebra $Q(3)$, 
\begin{eqnarray}
&[A,B]&=C.\label{k4}
\\&[A,C]&=2\{A,B\}-4aHQL_3+2c_1Q^3L_3+c_0QL_3+(c_2+c_3)B\nonumber\\&&\quad+a(c_2-c_3)H-\frac{1}{2}c_1(c_2-c_3)Q^2-\frac{1}{4}c_0(c_2-c_3)\label{k5},
\\&[B,C]&=-2B^2+8bAH+2a^2H^2-2(ac_1+2b)HQ^2-4bHL_3^2-4dAQ^2\nonumber\\&&\quad+\frac{1}{2}(c_1^2+4d)Q^4+2dQ^2L_3^2-4c_4A+(4b-ac_0)H\nonumber\\&&\quad+\frac{1}{2}(c_0 c_1+4c_4-4d)Q^2+2c_4L_3^2+\frac{1}{8}(c_0^2-16c_4)\label{k6}.
\end{eqnarray}
The Casimir operator of $Q(3)$ in terms of central elements is given by
\begin{eqnarray}
K&=&4a^2H^2L_3^2+4a^2H^2Q^2+a^2(c_2+c_3)H^2-4ac_1HQ^4-4(2b+ac_1)HQ^2L_3^2\nonumber\\&&-(2ac_0-2bc_2+ac_1c_2-2bc_3+ac_1c_3)HQ^2-2(ac_0-bc_2-bc_3)HL_3^2\nonumber\\&&-\frac{1}{2}(4bc_2+ac_0c_2+4bc_3+ac_0c_3-4bc_2c_3)H+4b(c_2-c_3)HQL_3+c_1^2Q^6 \nonumber\\&& +(c_1^2+4d)Q^4L_3^2+\frac{1}{4}(4c_0c_1 +c_1^2c_2 +c_1^2c_3 -4c_2d -4c_3d)Q^4+(c_0c_1+4c_4\nonumber\\&&-c_2d-c_3d)Q^2L_3^2-2d(c_2-c_3)L_3Q^3 +\frac{1}{4}(c_0^2+c_0c_1c_2+c_0c_1c_3-4c_2c_4\nonumber\\&&-4c_3c_4+4c_2d+4c_3d-4c_2c_3d)Q^2-2c_4(c_2-c_3)QL_3+\frac{1}{4}(c_0^2-4c_2c_4\nonumber\\&&-4c_3c_4)L_3^2 +\frac{1}{16}(c_0^2c_2+c_0^2c_3+16c_2c_4+16c_3c_4-16c_2c_3c_4)\label{k7}.
\end{eqnarray}
This is a main step in the application of the deformed oscillator algebra approach which relies on the quadratic algebra $Q(3)$ and the Casimir operator $K$. In order to derive the spectrum of the system we realize the quadratic algebra $Q(3)$ in terms of deformed oscillator algebras \cite{Das1, Das3} $\{\aleph, b^{\dagger}, b\}$ of the form
\begin{eqnarray}
[\aleph,b^{\dagger}]=b^{\dagger},\quad [\aleph,b]=-b,\quad bb^{\dagger}=\Phi (\aleph+1),\quad b^{\dagger} b=\Phi(\aleph).\label{kpfh}
\end{eqnarray}
Here $\aleph $ is the number operator and $\Phi(x)$ is well behaved real function satisfying 
\begin{eqnarray}
\Phi(0)=0, \quad \Phi(x)>0, \quad \forall x>0.\label{kpbc}
\end{eqnarray}
It is non-trivial to obtain such a realization and find the structure function $\Phi(x)$. The realization of $Q(3)$ is of the form $A=A(\aleph)$, $B=b(\aleph)+b^{\dagger}\rho(\aleph)+\rho(\aleph)b$, where
\begin{eqnarray}
A(\aleph)&=&\left\{(\aleph+u)^2-\frac{1}{4}-\frac{c_2+c_3}{4}\right\},
\\
b(\aleph)&=&\left[4aHQL_3-2c_1Q^3L_3-c_0QL_3-a(c_2-c_3)H+\frac{1}{2}c_1(c_2-c_3)Q^2\right.\nonumber\\&&\left.+\frac{1}{4}c_0(c_2-c_3)\frac{1}{4}\right]\frac{1}{4(\aleph+u)^2-1} ,
\\
\rho(\aleph)&=&\frac{1}{3. 2^{20}(\aleph+u)(1+\aleph+u)\{1+2(\aleph+u)^2\}},
\end{eqnarray}
and $u$ is a constant to be determined from constraints on the structure function.
\subsection{Unirreps and energy spectrum} 
We now construct the structure function $\Phi(x)$ from the realizations of the quadratic algebra ((\ref{k4})-(\ref{k6})) and the Casimir operator (\ref{k7}) as follows
\begin{eqnarray}
\Phi(x;u, H)&=&12288 \left[c_0^2 + c_0 (-8aH + 4 c_1 Q^2) + 4 [2 H \{2 a^2 H + b(1 - 2 (x+u))^2\}\right.\nonumber \\&&\left. - c_4 (1 - 2 (x+u))^2- \{4 a c_1 H + d(1 - 2 (x+u))^2\} Q^2 + c_1^2 Q^4]\right]\nonumber\\&&\times\left[c_2^2 -2 c_2 \{c_3 + (1 - 2 (x+u))^2 + 4 L_3 Q\}+ \{c_3 - (-1 + 2L_3 \right. \nonumber\\&&\left.+ 2 (x+u)) (-1 + 2 (x+u) - 2 Q)\} \{c_3 + (1 + 2 L_3 - 2 (x+u)) \right. \nonumber\\&&\left.\times(-1 + 2 (x+u) + 2 Q)\}\right].
\end{eqnarray}
We need to use an appropriate Fock space to obtain finite dimensional unirreps. Thus the action of the structure function on the Fock basis $|n, E\rangle$ with $\aleph|n, E\rangle =n|n,E\rangle$ and using the eigenvalues of $H$, $Q$ and $L_3$, the structure function becomes the following factorized form:
\begin{eqnarray}
\Phi(x;u, E)&=&-3145728 (c_4 - 2 bE + d q_2^2)[x+u-\frac{1}{2}(1-m_1+m_2)]\nonumber\\&&\times[x+u-\frac{1}{2}(1+m_1-m_2)][x+u-\frac{1}{2}(1-m_1-m_2)]\nonumber\\&&\times [x+u-\frac{1}{2}(1+m_1+m_2)]\left[x+u-\left(\frac{1}{2}+\frac{c_0-4aE+2c_1 q_2^2}{4\sqrt{c_4-2bE+dq_2^2}}\right)\right]\nonumber\\&&\times \left[x+u-\left(\frac{1}{2}-\frac{c_0-4aE+2c_1 q_2^2}{4\sqrt{c_4-2bE+dq_2^2}}\right)\right],\label{stf1}
\end{eqnarray}
where $m_1^2=c_2+(q_1-q_2)^2$, $m_2^2=c_3+(q_1+q_2)^2$, $q_1$  and $q_2$ are the eigenvalues of $L_3$ and $Q$ respectively.

For finite dimensional unitary representations we should impose the following constraints on the structure function (\ref{stf1}), 
\begin{equation}
\Phi(p+1; u,E)=0,\quad \Phi(0;u,E)=0,\quad \Phi(x)>0,\quad \forall x>0,\label{pro2}
\end{equation}
where $p$ is a positive integer. These constraints give $(p+1)$-dimensional unitary representations and their solution gives the energy $E$ and the arbitrary constant $u$. Thus all the possible structure function and energy spectra for $
\epsilon_1=\pm 1$, $\epsilon_2=\pm 1$,
\\Set-1:
\begin{eqnarray}
u=\frac{1}{2}(1 +\epsilon_1 m_1 +\epsilon_2 m_2), \quad \frac{2aE-c_1 q^2_2-\frac{c_0}{2}}{\sqrt{c_4-2bE+dq_2^2}}= 2 + 2 p  + \epsilon_1 m_1 + \epsilon_2 m_2 ,  \nonumber\\&&
\end{eqnarray}
\begin{eqnarray}
\Phi(x)&=&3145728 x(x +\epsilon_1 m_1)(x +\epsilon_2 m_2)(x+\epsilon_1 m_1+\epsilon_2 m_2)(1+p-x)\nonumber\\&&\times(1+ p+x +\epsilon_1 m_1+\epsilon_2 m_2 )s_1^2.
\end{eqnarray}
\\Set-2:
\begin{eqnarray}
&&u=\frac{1}{2}-\frac {2aE-c_1 q^2_2-\frac{c_0}{2}}{2\sqrt{c_4-2bE+dq_2^2}}, \quad \frac{2aE-c_1 q^2_2-\frac{c_0}{2}}{\sqrt{c_4-2bE+dq_2^2}}= 2 + 2 p  + \epsilon_1 m_1 + \epsilon_2 m_2,   \nonumber\\&&
\end{eqnarray}
\begin{eqnarray}
\Phi(x)&=& \frac{1572864 (1 + p)}{a}[x-1-p][1 + p- x+\epsilon_1 m_1 ][1+p-x+\epsilon_2 m_2 ]\nonumber\\&& \times[1+ p- x+\epsilon_1 m_1 +\epsilon_2 m_2 ][2+ 2 p- x+\epsilon_1 m_1 +\epsilon_2 m_2 ]\left[-2a(c_4 + d q_2^2)\right.\nonumber\\&&\left. + b\{c_0 + 2c_1 q_2^2 + 2 (2+2 p+\epsilon_1 m_1 +\epsilon_2 m_2 )s_1\}\right],
\end{eqnarray}
\begin{eqnarray*}
\text{where} &s^2_1&=c_4+d q_2^2+\frac{b}{2a^2}\left[b \eta_1^2-a c_0 - 2 a c_1 q_2^2\right.\\&&\left.\quad +\sqrt{\eta_1^2 \{b^2 \eta_1^2 - 2 a b (c_0 + 2 c_1 q_2^2) + 4 a^2 (c+4 + d q_2^2)\}}\right], \\& \eta_1&=2p + 2 + \epsilon_1 m_1 +\epsilon_2 m_2.
\end{eqnarray*}
The Structure functions are positive for the constraints $\varepsilon_1=1$, $\varepsilon_2=1$ and $m_1, m_2>0$. Using formula (\ref{dt1}) and $\nu_2=q_2$, we can write $m_1=\sqrt{c_2+(\nu_1-2\nu_2)^2}-\nu_1$ and $m_2=\sqrt{c_3+\nu_1^2}-\nu_1$. Making the identification $p=n_1+n_2$, the energy spectrum coincides with the physical spectra (\ref{en3}). The physical wave functions involve other quantum numbers and we have in fact degeneracy of $p+1$ only when these other quantum numbers would be fixed. The total number of degeneracies may be calculated by taking into account the further constraints on these quantum numbers.

\section{Conclusion}
In this paper, we have introduced a new superintegrable monopole system in the Taub-NUT space whose wave functions are given in terms of a product of  Laguerre and Jacobi polynomials. By construction, algebraically independent integrals of motion of the Hamiltonian (\ref{k1}) makes it a superintegrable system with monopole interactions. We have presented the quadratic algebra and corresponding Casimir operator generated by the integrals. The realization of this algebra in terms of deformed oscillator enable us to provide the finite dimensional unitary representations and the degeneracy of the energy spectrum of the monopole system.

{\bf Acknowledgements:}
The research of FH was supported by International Postgraduate Research Scholarship and Australian Postgraduate Award. IM was supported by the Australian Research Council through a Discovery Early Career Researcher Award DE 130101067. YZZ was partially supported by the Australian Research Council, Discovery Project DP 140101492. He would like to thank the Institute of Theoretical Physics, Chinese Academy of Sciences for hospitality and support.

\end{document}